# CAN CONFORMAL WEYL GRAVITY BE CONSIDERED A VIABLE COSMOLOGICAL THEORY? [1]

D. Elizondo and G. Yepes
Dep. de Física Teórica C-XI, Universidad Autónoma de Madrid, Cantoblanco 28049 Madrid SP

## ABSTRACT

We present exact solutions to the conformal Weyl gravity cosmological equations valid for both the matter and radiation dominated eras. We show that the cosmological models derived from this theory are not likely to reproduce the observational properties of our Universe. They fail to fulfill simultaneously the observational constraints on present cosmological parameters and on primordial light element abundances.

*Subject headings:* cosmology: theory – early universe – gravitation – nucleosynthesis

## 1. Introduction

Conformal Weyl gravitational theory is based on the assumption of local conformal invariance of the space-time geometry. It has been proposed as an alternative gravitation theory (c.f. Adler 1982; Zee 1983), for the very high energy limit. Many attemps have been made at trying to break the conformal invarince dynamically so as to obtain Einstein gravity as a phenomenological theory. Unfortunately, no attempt has been successful as yet. Quantum fluctuations bring back the Hilbert-Einstein action with a repulsive rather than atractive gravity, when breaking the conformal symmetry dynamically.

An alternative aproach to the symmetry breaking process has been recently proposed. Kazanas & Mannheim, in a series of papers, have suggested that Weyl gravity should be considered a theory in its own right, instead of looking for a low energy Einstein limit. They have found exact vacuum solutions (Mannheim & Kazanas 1989; 1990) and more recently, exact cosmological models, valid only for a radiation dominated universe (Mannheim 1992).

Two important properties of this theory were found: both the cosmological constant problem and the flatness problem are solved naturally in the context of conformal gravity (Mannheim 1990; 1992) and the Newtonian limit of the theory gives rise to linear terms that could explain the flat rotation curves of galaxies without the assumption of a massive dark matter halo (Mannheim & Kazanas 1989). They also showed that the Schwarzschild solution can be obtained at length scales where the linear terms are unimportant (i.e. $\ll$ galaxy size), therefore fulfilling the same weak field observational tests that Einstein's theory does.

Conformal Weyl gravity can be then considered as a valid theory of gravitation, at least in the weak field approximation. It is nevertheless necessary for the theory to be realistic to study the cosmological models and confront their predictions with the standard big bang cosmology. Mannheim (1992) has derived cosmological field equations for the Robertson-Walker (RW) metric. He was able to obtain analytical solutions for a radiation dominated universe. The main result is that all possible models have an hyperbolic geometry ($k < 0$) but, contrary to the Friedman-Robertson-Walker (FRW) models, with oscillatory and singularity-free dynamics. Exact analytical solutions describing the whole thermal history of the Universe were not quoted. Besides, early universe nucleosynthesis in these models has not been studied as yet, although it is well known that Primordial Nucleosynthesis (PN) is one of the best ways to test the viability of alternative metric theories of gravitation (Will 1981; Domínguez-Tenreiro & Yepes 1987; Serna, Domínguez-Tenreiro & Yepes 1992).

[1]To appear in *The Astrophysical Journal*, June 1994

In this letter we present an exact analytical solution of the Conformal Weyl gravity cosmological field equations. Mannheim's solution for the radiation dominated era is just an approximation to this general one. We study those models that are compatible with present cosmological parameter values and compute the PN of these models. A comparison with the estimated primordial abundances of light elements shows that the models do not pass the PN test.

In what follows we shall use units such that $c = \hbar = k_B = 1$.

## 2. Cosmological equations

Conformal Weyl gravity is described by the action

$$I_W = -\alpha \int C_{\lambda\mu\nu\kappa} C^{\lambda\mu\nu\kappa} \sqrt{-g} d^4x \tag{1}$$

where $C_{\lambda\mu\nu\kappa}$ is the Weyl tensor and $\alpha$ a dimensionless coefficient. The action for the matter content of the theory, $I_M$, can be of the form (Mannheim 1992)

$$I_M = -\int [S^\mu S_\mu/2 + \lambda S^4 - S^2 R^\mu_\mu/12 + i\overline{\psi}\gamma^\mu(x)(\partial_\mu + \Gamma_\mu(x))\psi - hS\overline{\psi}\psi]\sqrt{-g} d^4x, \tag{2}$$

where $\psi(x)$ is a fermion field representing all matter and $S(x)$ is a conformal scalar field that yields symmetry breaking and renders the particles massive. $R^\mu_\mu$ is the scalar curvature, $\Gamma_\mu(x)$ the fermion spin connection and h and $\lambda$ are dimensionless coupling constants.

Due to the conformal invariance of the theory the energy momentum tensor $T_{\mu\nu}$ is kinematically traceless. Assuming a Robertson-Walker metric and the tracelessness of $T_{\mu\nu}$ the following set of field equations is derived (Mannheim 1992)

$$\rho + \lambda S^4 + S^2(\dot{R}^2 + k)/2R^2 = 0 \tag{3}$$

$$3p - \rho - 4\lambda S^4 - S^2(R\ddot{R} + \dot{R}^2 + k)/R^2 = 0 \tag{4}$$

with $\rho$ and $p$ being the total density and pressure of a perfect fluid and $R$ the expansion factor of the RW metric. Conformal invariance implies that $S(x) \equiv S$ is a spacetime independent constant. In addition, we also have the energy conservation equation for a perfect fluid

$$3p + 3\rho + R\frac{d\rho}{dR} = 0 \tag{5}$$

## 3. Exact solutions

From the above set of equations we have been able to obtain exact solutions for the dynamical variables as a function of the temperature. From equations (3) and (4) and making use of the Hubble parameter $H \equiv \dot{R}/R$ we get

$$\dot{H} = \frac{3p}{S^2} + \frac{\rho}{S^2} - 2\lambda S^2 - H^2 \tag{6}$$

Introducing the definition of the decceleration parameter, $q \equiv -\ddot{R}R/\dot{R}^2 = -(\dot{H}/H^2 + 1)$, in the above equation

$$2\lambda S^2 = \frac{3p + \rho}{S^2} + qH^2 \tag{7}$$

This is a spacetime independent equation, therefore we can obtain the value of $\lambda S^2$ evaluating eq. (7) at the current epoch.

$$2\lambda S^2 = \frac{3p_0 + \rho_0}{S^2} + q_0 H_0^2 \tag{8}$$

In what follows, a "0" subscript will stand for current parameter values. We are thus left with only one free parameter, the value of $S$.

During most of the thermal evolution of the Universe (except in the particle-antiparticle annihilation era) the comoving entropy is conserved ($RT$ =cte). In this case, eq (5) leads to (e.g. Weinberg 1972)

$$\frac{dT}{dt} = -HT \tag{9}$$

that, in turn, can be used to transform the dependence of eq. (6) from time to temperature. Furthermore we can put $\rho = \rho^M + \rho^R$, where $\rho^M = \rho_0^M (T/T_0)^3$ is the matter density and $\rho^R = \rho_0^R (T/T_0)^4$ is the density of the relativistic fluid component. Also, $p = p^M + p^R$, with $p^M = 0$ and $p^R = \rho^R/3$.

Changing to dimensionless variables $\mathcal{H} \equiv H/H_0$ and $\mathcal{T} \equiv T/T_0$, and making use of eq. (7) and (9), we obtain

$$\frac{d(\ln \mathcal{H})}{d(\ln \mathcal{T})} = 1 + \frac{q_0}{\mathcal{H}^2} + \frac{3\Omega_0^M}{8\pi G S^2 \mathcal{H}^2}\left[1 - \mathcal{T}^3 + 2\frac{\rho_0^R}{\rho_0^M}(1 - \mathcal{T}^4)\right], \tag{10}$$

where we have introduced the density parameter $\Omega_0^M = 8\pi G \rho_0^M / 3H_0^2$. The solution of the above differential equation gives the temperature dependence of H. After a straightforward calculation we get,

$$\mathcal{H} = \left[-2B\mathcal{T}^4 - 2A\mathcal{T}^3 + (1 + q_0 + 3A + 4B)\mathcal{T}^2 - (q_0 + A + 2B)\right]^{1/2}, \tag{11}$$

with $A = 3\Omega_0^M / 8\pi G S^2$ and $B = 3\Omega_0^R / 8\pi G S^2$. It defines a set of cosmological models with only one free dimensionless parameter, $GS^2$. It must be noted that, contrary to the FRW cosmology, in this case $q_0$ is not directly related with $\Omega_0$.

From eqs. (3) and (4) we get:

$$\dot{H} - \frac{k}{R^2} - \frac{3(\rho + p)}{S^2} = 0, \tag{12}$$

substituting $\dot{H}$ in the above equation, by means of eq. (6) and eq. (7) we obtain:

$$\frac{k}{R^2} = -H^2 - \frac{2\rho}{S^2} - \frac{3p_0 + q_0}{S^2} - q_0 {H_0}^2. \tag{13}$$

If we make use of the solution (11), after some arrangements we derive:

$$R = \frac{-k}{H_0 \mathcal{T}(1 + q_0 + 3A + 4B)^{1/2}} \tag{14}$$

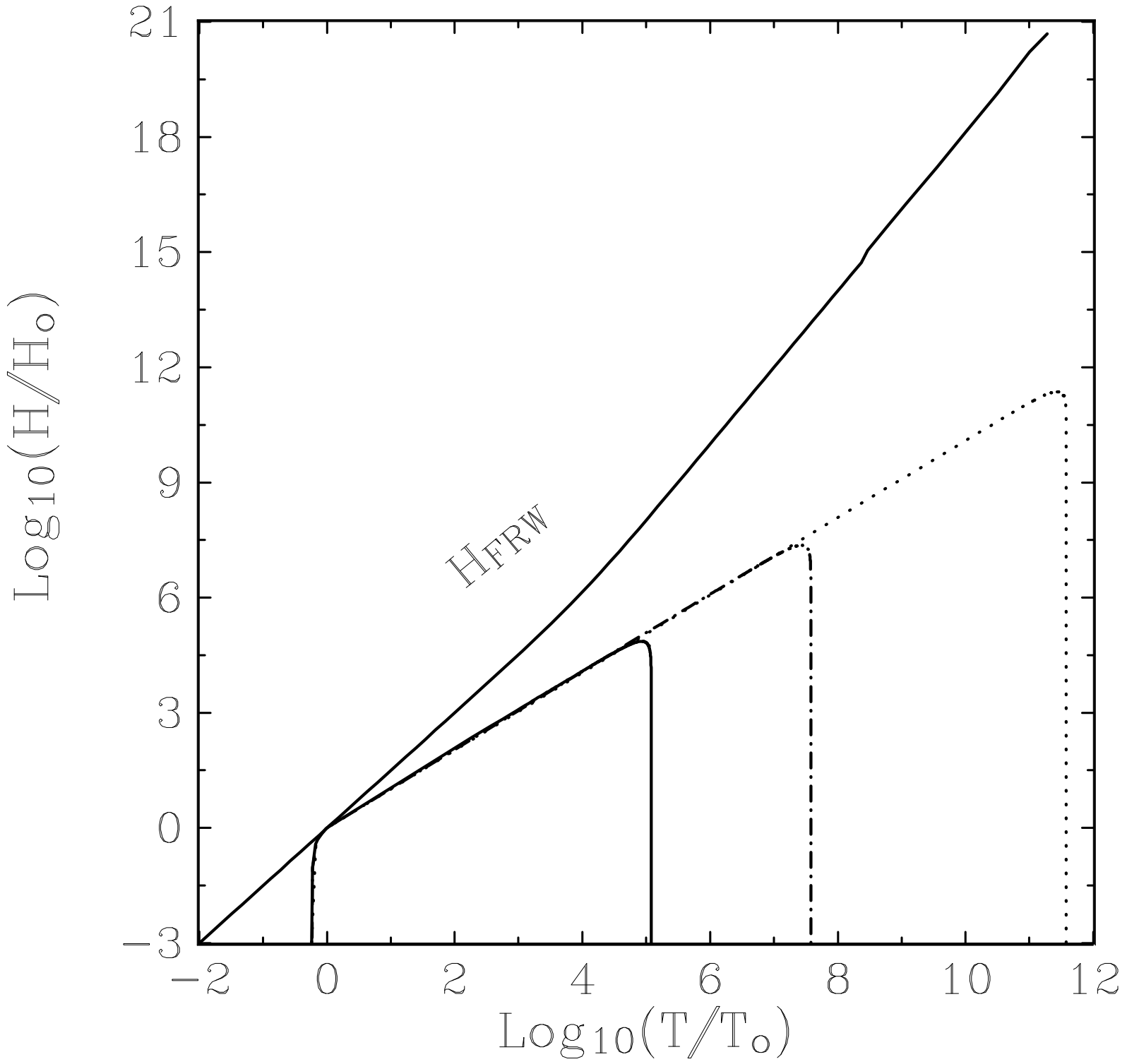

Fig. 1.— The Hubble parameter as a function of temperature for $q_0 = 0.5$, $\Omega_0 = 1$ and three different $GS^2$ values: $10^5$ (solid line), $10^{10}$ (dotted-dashed line) and $10^{18}$ (dotted line). For comparison, we also plot $\mathcal{H}(\mathcal{T})$ for the standard euclidean FRW model.

which shows that only open ($k < 0$) universe models are possible since we know that $H_0 > 0$. This result is directly related with the fact that the dimensionless constant $\lambda$ must be positive in order to have a well defined matter action (see. eq. (2)). From eq. (3) it can be seen that $\lambda > 0$ implies $k < 0$, since $\rho$ is a positive quantity.

There is not a general analytical solution for $R(t)$. Nevetherless it can be easily seen that for a radiation dominated universe ($A \sim 0$) we can derive, from eq. (9) and (11), the same oscillatory solution found by Mannheim (1992).

In our case, if we want to reach high enough temperatures for the PN to take place, we need that $\mathcal{H}^2 > 0$, at least for $\mathcal{T} \leq 10^{12}$. This can be achieved only if $A$ and $B$ in eq. (11) are very small. This is equivalent to say that $GS^2$ must be very large. To show this behavior, we have plotted in Fig. 1 the exact solution $\mathcal{H}(\mathcal{T})$, given by eq. (11), for a wide range of $GS^2$ values and for the special case with $q_0 = 0.5$ and $\Omega_0 = 1$. The dependence of $\mathcal{H}(\mathcal{T})$ on the value of the cosmological parameters is negligible (if we restrict ourself to their observational limits) because the dynamical behavior is governed by the value of the scalar field $S$.

From Fig. 1 we see that there always exists a maximum and minimum temperature corresponding to the minimum and maximum radius of the universe model given by eq. (14).

We can get the temperature at which $H$ reaches the maximum value by finding the maxima of eq. (11). It is

$$\mathcal{T}_{Hmax} = \left[1 + \frac{9A^2}{64B^2} + \frac{1}{4B}(1 + q_0 + 3A)\right]^{1/2} - \frac{3A}{8B} \tag{15}$$

For high $GS^2$ values, $\mathcal{T}^2_{Hmax} \sim GS^2(\Omega_0^R)^{-1}$. In order to have temperatures $\geq 10^9$K for the PN to take place, we should need $GS^2 \geq 10^{14}$. For these values, $\mathcal{H}$ will be approximated by the asymptotic form

$$\mathcal{H}_{asym} = \sqrt{(1 + q_0)\mathcal{T}^2 - q_0} \tag{16}$$

We see the manifest linearity of $\mathcal{H}(\mathcal{T})$, contrary to the FRW model where $\mathcal{H} \propto \mathcal{T}^2$ (see Fig 1). Thus, the ratio between the expansion factor of the standard FRW model and the one obtained here will be

$$\xi \equiv \frac{\mathcal{H}_{FRW}}{\mathcal{H}} \propto \mathcal{T} \tag{17}$$

This will be crucial for the viability of the models, since the PN abundances are very sensitive to large desviations from the $H_{FRW}$ values at PN temperatures (Domínguez-Tenreiro & Yepes 1987).

We have nevertheless computed the light element production in these models. We have used an updated version of the original Wagoner (1969) PN code that has been developped at Fermilab (Kawano 1992). The code has been modified in order to account for the particular dynamics of the models described by this theory. To integrate the set of nuclear kinetic equations it is necessary to know how fast the universe cools. This is given by $\mathcal{H}(\mathcal{T})$. But the analytical solution (11) is not valid during the $e^+e^-$ annihilation phase, because the assumption $RT = cte$ is no longer valid. It was necessary to numerically compute $H(T)$ by eq. (13), using eq. (11) to set up the initial conditions at lower temperatures, where the analytical solution is valid.

Table 1: Light element production in conformal Weyl gravity cosmological models.

| $h$ | $q_0$ | $\Omega_0 h^2$ | $\Omega_0^B h^2$ | $D/H$ | $(D +^3 He)/H$ | $Y_{^4He}$ | $^7Li/H$ |
|---|---|---|---|---|---|---|---|
| 0.5 | 0.5 | 0.25 | 0.0075 | $1.19 \times 10^{-20}$ | $3.87 \times 10^{-8}$ | $2.88 \times 10^{-4}$ | $6.21 \times 10^{-9}$ |
| . | . | . | 0.025 | $3.13 \times 10^{-20}$ | $1.51 \times 10^{-8}$ | $1.73 \times 10^{-3}$ | $1.53 \times 10^{-8}$ |
| . | . | . | 0.25 | $1.13 \times 10^{-19}$ | $4.44 \times 10^{-18}$ | $6.17 \times 10^{-1}$ | $2.05 \times 10^{-13}$ |
| . | . | 2.5 | 0.0075 | $1.42 \times 10^{-20}$ | $5.97 \times 10^{-8}$ | $8.60 \times 10^{-4}$ | $2.85 \times 10^{-9}$ |
| . | . | . | 0.25 | $1.40 \times 10^{-19}$ | $9.48 \times 10^{-18}$ | $1.04 \times 10^{-1}$ | $1.29 \times 10^{-11}$ |
| . | . | . | 0.5 | $2.36 \times 10^{-19}$ | $1.01 \times 10^{-17}$ | $2.23 \times 10^{-1}$ | $4.06 \times 10^{-13}$ |
| . | . | . | 1.25 | $5.69 \times 10^{-19}$ | $2.11 \times 10^{-17}$ | $1.55 \times 10^{-1}$ | $1.07 \times 10^{-13}$ |
| . | . | . | 2.5 | $1.27 \times 10^{-18}$ | $6.52 \times 10^{-17}$ | $2.07 \times 10^{-2}$ | $2.01 \times 10^{-12}$ |
| . | 2 | 0.25 | 0.0075 | $7.08 \times 10^{-21}$ | $3.63 \times 10^{-8}$ | $3.04 \times 10^{-3}$ | $7.32 \times 10^{-9}$ |
| . | . | 2.5 | 2.5 | $7.49 \times 10^{-19}$ | $2.80 \times 10^{-17}$ | $5.98 \times 10^{-2}$ | $4.74 \times 10^{-13}$ |
| 1 | 0.5 | 0.25 | 0.0075 | $6.38 \times 10^{-21}$ | $6.18 \times 10^{-8}$ | $1.89 \times 10^{-3}$ | $3.00 \times 10^{-9}$ |
| . | . | 2.5 | 2.5 | $5.24 \times 10^{-19}$ | $1.36 \times 10^{-17}$ | $2.01 \times 10^{-1}$ | $5.31 \times 10^{-14}$ |

Note. — Abundances are by number relative to Hydrogen, except for $^4He$ that is given as mass fraction. $h = H_0/100$ km s$^{-1}$ Mpc$^{-1}$)

In Table 1 we present the light element abundances for a wide range of baryonic content: from the lower observational bound $\Omega_0^B h^2 = 0.0075$ (Weinberg 1972) up to $\Omega_0^B h^2 = 2.5$ . This is equivalent to a baryon-to-photon ratio, $\eta_B$, ranging from $\eta_B = 2.05 \times 10^{-10}$ up to $6.85 \times 10^{-8}$ .

There was no any combination of the parameter values that could give light element abundances within the observational limits (Boesgaard & Steigman 1985; Smith, Kawano & Malaney 1993). The PN process depends basically on the expansion rate and on the baryonic content. As we said before, the dynamical evolution of these models is not very much affected by the value of the cosmological parameters. Therefore, as can be seen in Table 1, the PN process is not very sensitive to the particular values of $H_0, q_0$, or $\Omega_0$ .

A correct $^4He$ abundance can be obtained in models with big amounts of baryonic matter ($\Omega_0^B h^2 \sim 0.5 - 2.5$). The $^7Li$ abundance is also in agreement for a wide range of parameter values. But the main drawback of these models is the deuterium abundance. Because the universe described by these models expands much slower than in the standard FRW models, (see. eq. (17)), the nuclear reactions that burn the deuterium nuclei have more time to act consuming practically all of it. At the end of the PN process almost nothing of deuterium is left in contradiction with the current observational bounds that give a lower limit for the primordial abundance of $D/H \geq 1.8 \times 10^{-5}$ (Smith, Kawano, & Malaney 1993).

## 4. Conclusions

We have obtained exact analytical solutions to the conformal Weyl gravity cosmological equations. We show that previous solutions, valid only for the radiation dominated era are only approximations to this solution. The main characteristic of the models are that they have an hyperbolic geometry but with an oscillatory behavior. Temperature is bounded between a maximum and a minimum value, with no singularity state. In order to have high enough temperature to allow PN to take place, we need to assume a very high value for the dimensionless parameter $GS^2$.

PN abundances in these models are incompatible with observational constraints for all range of parameter values. We conclude that conformal gravity cosmological models are very unlikely to give a realistic description of our Universe.

We acknowledge M. Smith for kindly sending us a copy of the NUC123 PN code. We would also like to thank C.P. Martin for very usefull discussions.

---

This preprint was prepared with the AAS LaTeX macros v3.0.